\documentclass{aa}
\input{psfig.sty}
\catcode`\@=11 
\def\gs{\ifmmode{\mathrel{\mathpalette\@versim>}}
\else{$\mathrel{\mathpalette\@versim>}$}\fi}
\def\ls{\ifmmode{\mathrel{\mathpalette\@versim<}}
\else{$\mathrel{\mathpalette\@versim<}$}\fi} 
\def\@versim#1#2{\lower2.9truept \vbox{\baselineskip 0pt \lineskip 0.5truept
\ialign{$\m@th#1\hfil##\hfil$\crcr#2\crcr\sim\crcr}}} 
\catcode`\@=12

\def\kms{\,km\,s$^{-1}$}

\def\es{ergs s$^{-1} \ $}
\def\cts{counts s$^{-1} \ $}

\def\kms{km s$^{-1}$}

\def\lae{\mathrel{<\kern-1.0em\lower0.9ex\hbox{$\sim$}}}
\def\gae{\mathrel{>\kern-1.0em\lower0.9ex\hbox{$\sim$}}}

\begin{document}

\title{Unveiling the AGN powering the ``Composite" Seyfert/Star-forming 
galaxy NGC 7679: BeppoSAX and ASCA results}

\author{R. Della Ceca \inst{1}, S. Pellegrini \inst{2}, L. Bassani \inst{3}, 
V. Beckmann \inst{4}, M. Cappi \inst{3},
 G.G.C. Palumbo \inst{2,5}, G. Trinchieri \inst{1}, A. Wolter \inst{1}
       }

\offprints{R. Della Ceca}

\institute{
Osservatorio Astronomico di Brera, via Brera 28, I-20121, Milan, Italy.
\and
Universit\`a di Bologna, Dipartimento di Astronomia, via Ranzani 1, I-40127,
Bologna, Italy.
\and
Istituto TeSRE/CNR Via Gobetti 101, I-40129, Bologna, Italy.
\and 
INTEGRAL Science Data Centre, Chemin d'Ecogia 16, CH-1290, Geneva, Switzerland.
\and
Agenzia Spaziale Italiana (ASI), Viale Liegi 26, I-00198, Rome, Italy
}

\date{Received 19 February 2001/Accepted 5 June 2001}

  \abstract{We discuss {\it Beppo}SAX observations and archive ASCA
   data of NGC 7679, a nearby,  nearly face-on SB0 galaxy in
   which starburst and AGN activities coexist.  
   The X-ray
   observations reveal a bright ($L_{0.1-50 keV} \sim 2.9 \times
   10^{43}$ erg s$^{-1}$) and variable source having a minimum
   observed doubling/halving time scale of $\sim 10 - 20$ ksec.  A simple
   power law with photon index of $\Gamma \sim 1.75$ and small
   absorption ($N_H < 4\times 10^{20}$ cm$^{-2}$) can reproduce the
   NGC 7679 spectrum 
   from 0.1 up to 50 keV.  
   These X-ray properties are unambiguous signs of Seyfert 1 activity in the nucleus of NGC
   7679.  The starburst activity, revealed by the IR emission, 
   optical spectroscopy and H$ \alpha$ imaging, and 
   dominating in the optical and IR bands, is clearly overwhelmed by the
   AGN in the X-ray band.  Although, at first glance, this is similar
   to what is observed in other starburst-AGN galaxies (e.g. NGC 6240,
   NGC 4945),
   most strikingly here and at odds with the
   above examples, the X-ray spectrum of NGC 7679 does not appear to be
   highly absorbed.  
   The main
   peculiarity of objects like NGC 7679 is not the strength of their
   starburst but the apparent optical weakness of the Seyfert 1
   nucleus when compared with its X-ray luminosity.  To date NGC 7679
   is one of the few Seyfert 1/ Starburst composites for which the
   broad-band X-ray properties have been investigated in detail.  The
   results presented here imply that optical and infrared spectroscopy
   could be highly inefficient in revealing the presence of an AGN in
   these kinds of objects, which instead is clearly revealed from
   X-ray spectroscopic and variability investigations.
    }

\authorrunning{Della Ceca et al.,}
\titlerunning{{\it Beppo}SAX and ASCA results on NGC 7679}
\maketitle

\section{Introduction}

The connection between starburst and AGN activity is one of the most
discussed and debated topics of modern cosmology.  This connection has
been used to explain the overall properties and the physical
nature of the Ultraluminous Infrared Galaxies (ULIRG, i.e. sources
with bolometric luminosities in excess of $10^{12}$ $L_{\odot}$, mostly
emitted in the mid and far infrared), and recently (Fabian et al.
1998) as an elegant way to produce absorption in 
low luminosity ($L_x \sim 10^{43}$
\es) AGNs. 
Therefore, understanding this phenomenon is clearly of high priority 
(see Veilleux, 2000 for a recent review).
    
X-ray spectroscopy is a very powerful tool to shed light on the starburst-AGN 
connection, mainly because X-rays are emitted  closer to the
primary emitting source than optical emission lines, which represent
reprocessed radiation from interstellar gas (e.g., Terashima et al. 2000).
In addition, while the optical line
diagnostic tools, based on emission line intensity ratios, are often compatible
with both starburst and AGN activity (Terlevich et al. 1992), the X-ray
properties of starbursts (i.e. mostly thermal emission) 
and AGNs (i.e. mostly non-thermal power-law emission) 
are quite different.
In ``pure" starburst galaxies, X-rays likely  
derive from stellar objects (type II supernovae, O stars, high and low
mass X-ray binaries) as well
as galactic winds and superwinds (see Heckman, 2000 and references therein). 
X-ray observations of
starburst galaxies (see e.g. Della Ceca et al. 1997; Dahlem et
al. 1998; Ptak et al. 1999; Cappi et al., 1999;  Pietsch et al., 2001 and 
Griffiths et al., 2000)
revealed very
complex spectra, which required the presence of at least two or three
thermal emission components with temperatures up to a few keV.
If coming from ``pure" AGN-like activity rather than from a starburst, 
the nuclear X-ray
emission is expected to be stronger, more compact, variable, harder and
described, to a first approximation, by a power law model  with photon index 
$\Gamma \simeq 1.9 \pm 0.15$ (e.g., Nandra et al. 1997);
iron-K fluorescence emission lines with a mean equivalent width 
(EW) of $\sim 230$ eV are also often present
(e.g., Nandra et al. 1997). 
If the nuclear emission intercepts absorbing
material (e.g. a molecular torus), 
absorption effects (low energy cutoff, strong 
iron-K fluorescence emission lines) are expected (see Bassani et al., 1999 
and references therein).
Starburst/Seyfert composites have been also found and studied in detail in the 
last years; it is now clear that heavy absorption 
($N_H > 10^{23}$ cm$^{-2}$)
in the nuclear region and a strong  circumnuclear starburst are responsible  
for the
broad-band  properties of most of these objects 
(see Levenson, Weaver and Heckman, 2001 and  
Iwasawa, 1999 and references therein);
their X-ray spectra are usually described by a 
combination of reflected/trasmitted AGN emission plus starburst thermal 
emission.
   
A case study to investigate the starburst-AGN connection 
is the barred, nearly face-on ($i \sim 30^o$)
SB0 galaxy NGC 7679 (also known as MKN 534 and Arp 216; z=0.0177, 
Kewley et al. 2001).
Whether an active nucleus is present in NGC 7679, and to what extent 
starburst and AGN activities power the optical emission
lines, or in general the total emission, is not clear yet (see Section 2). 
We present here results obtained from a $BeppoSAX$ observation 
and archive ASCA data which help us 
clarify the nature of this galaxy. 


\section{Why NGC 7679 ?}

NGC 7679 was originally selected by us for investigation of its 
starburst properties in the X-ray band.
In fact, the H$\alpha$ image reveals a luminous nuclear star-formation complex
resolved into bright clumps of emission and a bright circumnuclear
starburst region that is about 18$^{\prime \prime}$ 
($\sim 9.3$ kpc) in diameter
(Pogge \& Eskridge 1993).  NGC 7679 also shows quite a strong infrared
luminosity from IRAS data (log$L_{FIR}/L_{\odot}$=11.10), a ratio
$L_{FIR}/L_B \sim 1$, and infrared colors typical of a starburst
galaxy (Kewley et al. 2001).
Further support for the presence of a starburst in NGC 7679 are a) the existence
of a bar; b) signs of tidal distortion due to the interaction with a faint
companion $\sim 1^{\prime}$ east; and c) the physical association with the
Seyfert 2 galaxy NGC 7682 (that lies $\sim 4^{\prime}\hskip -0.1cm .5$ east), as
revealed by a stream of ionized gas connecting the two galaxies (Durret \&
Warin 1990).  These characteristics are expected to enhance the gas flow
towards the galaxy center and are thought to trigger a
starburst (see e.g. Combes, 2000 for a review of this topic).  

Optical spectroscopy  by Kewley et al. (2001) reveals  narrow optical 
emission lines (HII region-like) as well as a weak broad 
(FWHM $\sim$ 2000 \kms) H$\alpha$ component
\footnote { 
We also have performed optical spectroscopic observations of NGC 7679 on 
September
14, 1999 at the Mt. Orzale 152cm telescope of the
{\it Osservatorio Astronomico di  Bologna}.
The observed optical spectral properties (line features, FWHM  
and ratios) support the classification of NGC 7679 as a
starburst galaxy (cf. Veilleux and Osterbrock, 1987
and Ho, Filippenko and Sarget, 1997), 
in good agreement with the results obtained 
independently by Kewley et al. 2001. 
We do not detect the broad H$ \alpha$ component 
at the bottom of the strong and  narrow H$ \alpha$ line, but 
the S/N ratio of our spectra is probably not adequate for
this purpose.
}.
Therefore, both the broad-line region of a
central AGN and the narrow-line region dominated by the excitation of
a starburst could have been detected in the optical spectrum
\footnote {
A caveat, however, is in order. The FWHM ($\sim 2000$ km s$^{-1}$) and
luminosity ($\sim 8 \times 10^{40}$ erg s$^{-1}$, Keweley, private
communication) of the broad $H \alpha$ component detected in NGC 7679 could
in principle  be produced by a supernova explosion.  A clear example is  
the type II Supernova 1997ab found in the dwarf galaxy HS 0948+2018  (Hagen,
Engels and Reimers, 1997); 
this type II supernova has a broad (FWHM $\sim 2500$ km s$^{-1}$)
$H \alpha$ component with a luminosity  that  is
more than an order of magnitude higher than that of NGC 7679.
}.


Finally, NGC 7679 was previously imaged in X rays with the {\it Einstein}
Observatory IPC (890 s exposure, Fabbiano et al. 1992) and with the
ROSAT PSPC during the ROSAT All Sky Survey (RASS) for a 381 s exposure
(1RXS J232846.9+033042, www.xray.mpe.mpg.de/cgi-bin/rosat/rosat-survey).  
The {\it Einstein} IPC count rate ($2.7 \times 10^{-2}$ \cts), when
converted using a 5 keV thermal bremsstrahlung model filtered by
Galactic absorption, gave a (0.2--4) keV flux of
$1.1\times 10^{-12}$ erg cm$^{-2}$ s$^{-1}$ and a luminosity of
$1.5\times 10^{42}$ erg s$^{-1}$.  The ROSAT PSPC (0.1 - 2.4 keV)
count rate is $(6.75 \pm 1.5) \times 10^{-2}$ \cts, corresponding to an observed 
flux (unabsorbed luminosity) of $9\times 10^{-13}$ erg cm$^{-2}$ s$^{-1}$
($1.7\times 10^{42}$ erg s$^{-1}$).  
No X-ray spectral information was available in literature prior 
to the results discussed here.
Some of the main characteristics of NGC 7679 are reported in Table 1.

\begin{table*}
\caption[] { General characteristics of NGC 7679}
\begin{flushleft}
\begin{tabular}{ l  l  l  l l  l  l l  l l  l }
\noalign{\smallskip}
\hline
\noalign{\smallskip}
Type$^a$  & RA     & Dec    & d$\, ^b$ & Optical$^b$    & $B_{\rm T}^0 \, ^a$ 
& log$L_{\rm B} \, ^a$ & log$L_{\rm FIR} \, ^b$
& Size$^a$ &  $N_{\rm H,Gal} \, ^c$ \\
          &(J2000) &(J2000) &  (Mpc)   & classification &   (mag)           
& ($L_{\odot}$)  & ($L_{\odot}$)      
& (arcmin)   &  (cm$^{-2}$)           \\
\noalign{\smallskip}
\hline
\noalign{\smallskip}
 SB0 & $23^h 28^m 46^s \hskip -0.1truecm .8$ & $03^{\circ} 30^{\prime} 
41^{\prime\prime} $ & 107 & SB+AGN & 12.89 & 11.09 & 11.10 &
1.3$\times$0.9  & 4.54$\times 10^{20}$ \\
\noalign{\smallskip}
\hline
\end{tabular} 
\end{flushleft}
\bigskip

$^a$ from de Vaucouleurs et al. 1991.  $B_{\rm T}^0$ is the total B magnitude,
corrected for Galactic and internal extinction.
The total B band luminosity, $L_{\rm B}$, 
has been derived using the indicated distance and 
$B_{\rm T}^0$.

$^b$ From Kewley et al. (2001). Distance based on $H_0=50$ km s$^{-1}$ Mpc$^{-1}$, $q_o=0$ 
and z=0.0177; at this distance 1$^{\prime} = 31.13$ kpc.
The infrared luminosity has been adapted for $H_0=50$ km s$^{-1}$ Mpc$^{-1}$.

$^c$ Galactic neutral hydrogen column density along the line of sight
from the Leiden/Dwingeloo Survey (Hartman and Burton, 1997).

\end{table*}

\section {{\it Beppo}SAX Observations and Data Preparation}           

NGC 7679 was observed by $BeppoSAX$ on December 6, 1998 (the journal of
the observation and source count rates are given in Table 2). 
In this paper we use data from
three Narrow Field Instruments: the Low Energy Concentrator
Spectrometer (LECS, Parmar et al. 1997), the Medium Energy
Concentrator Spectrometer (MECS, Boella et al. 1997), and the Phoswich
Detector System (PDS, Frontera et al. 1997). HPGSPC data (Manzo et
al. 1997) are not considered since the source is too faint to be
detected.

The cleaned and linearized data produced by the {\it BeppoSAX} Science
Data Center 
(http://www.asdc.asi.it/bepposax/)
have been analyzed using  standard software
(XSELECT v1.4, FTOOLS v4.2 and XSPEC v10.0).
The PDS data reduction was performed using XAS (V2.0, Chiappetti and Dal Fiume
1997) and SAXDAS (v1.3.0) software packages and yielded consistent results.

NGC 7679 is detected with a high signal-to-noise ratio (S/N) 
in the LECS and MECS instruments (see Table 2) but is not 
spatially resolved; the center of the X-ray emission was found to be 
coincident, within the positional accuracy allowed by $BeppoSAX$ data, 
with its optical nucleus.
To maximize the statistics and the S/N, the
MECS and LECS source counts have been extracted from a circle of $4^{\prime}$
radius around the centroid of the X-ray emission.
Background counts were extracted from high Galactic latitude ``blank" 
fields (provided by the {\it Beppo}SAX Science Data Center) using an  
extraction region which corresponds in size
and detector position to that used for the source.

A very small elongation (at a few percent of the peak level) appears 
in the MECS image at
the position of the Seyfert 2 galaxy NGC 7682,  
which lies $4^{\prime}\hskip -0.1cm .5$ east of NGC 7679 (see  Section 2).
NGC 7682 is not detected
in the ROSAT All Sky Survey; from the original RASS data we
can evaluate a flux upper limit (0.1-2.4 keV)  
which is at least a factor of 10 fainter than that of NGC 7679.  
To investigate if this source could contaminate the spectral results on 
NGC 7679 we have excluded the MECS counts deriving from 
a circle of $2^{\prime}$
radius centered on the position of NGC 7682, obtaining essentially the
same results as those reported in Section 5.  

Another X-ray source is detected with S/N $\sim 10$ in the MECS image at about
$19^{\prime}$  to the south-west of NGC 7679 (RA = $23^h 27^m 35^s \hskip
-0.1truecm .8$, Dec = $03^{\circ} 22^{\prime} 50^{\prime\prime}$).
A RASS object (1RXS J232735.2+032334,  
www.xray.mpe.mpg.de/cgi-bin/rosat/rosat-survey) 
is positionally consistent with this 
source; the RASS count rate is $0.03 \pm 0.01$, corresponding to an
observed flux (0.07 - 2.4 keV) of $\sim 4\times  10^{-13}$  erg cm$^{-2}$ s$^{-1}$.
There are no cataloged NED or SIMBAD objects inside a circle of $2^{\prime}$
radius from the RASS position.

Finally, there is a $6 \sigma$ detection in the PDS, also after 
conservative subtraction of the systematics residuals (currently evaluated at
$\sim$0.02 counts s$^{-1}$ 
in the 13--200 keV band, Guainazzi and Matteuzzi 1997).
There are no cataloged and bright (2-10 keV) 
X-ray sources, other than NGC 7679, in the PDS field of view.

\begin{table*}
\caption[Table 2.]{ {\it Beppo}SAX Observation Log}
\begin{flushleft}
\begin{tabular}{llllllllllllll}
\noalign{\smallskip}
\hline
\noalign{\smallskip}
Sequence & Date          & \multicolumn{3}{l}{Exposure time (ks)}  &\  & 
\multicolumn{3}{l}{Count Rate$^a$ (counts s$^{-1}$)}                 \\
\cline{3-5}\cline{7-9}
 Number &     & LECS & MECS & PDS & & LECS & MECS & PDS   \\
        &     &      &      &     & & 0.12--4 keV & 1.65--10 keV & 15-50 keV \\
\noalign{\smallskip}
\hline
\noalign{\smallskip}
40631001 & 1998 Dec 6--9 & 39.8 & 90.9 & 45.3 & & 0.041$\pm 0.001$ & 
0.073$\pm 0.001$ & 0.119$\pm 0.017$     \\ 
\noalign{\smallskip}
\hline
\end{tabular} 
\end{flushleft}
\bigskip

$^a$ Background subtracted source count rates, with $1 \sigma$ photon counting 
statistics errors. Extraction radii for LECS and MECS are $4^{\prime}$.
The net counts from NGC 7679 represent about $92\%$ 
of the total gross counts in the LECS or MECS source region.
The PDS counts correspond to a $6\sigma$ detection. 
\end{table*}

\section{X-ray Variability}
 
The LECS and MECS light curves extending over the whole observing period are
shown in Fig. 1 (LECS) and 2 (MECS).
A flux increase of roughly a factor 2.4  is present between $\sim 9\times 10^4$
and $\sim 1.3\times 10^5$ s from the start of the observation in 
both  LECS and MECS data.
The flux decreases
afterwards, until the end of the observation, reducing itself by a factor of
three.  No equivalent variations in the background are 
found.  This trend is present in all the energy bands considered (i.e.,
0.1--2 and 2--4 keV for the LECS, and 1.65--4 and 4--10 keV for the MECS). 
The flux variability seems to be not associated with spectral variations 
(see ratios in figure 1 and 2) and, 
therefore, it has to be attributed to an overall continuum variation; 
this is confirmed by the spectral analysis reported in
Sect. 5.2. From the flux change rate of the light curves 
we estimate a minimum doubling/halving time in the
range 10--20 ksec. Variability on smaller timescales cannot be
investigated due to the limited statistics available.

\begin{figure}[htb]
\parbox{10cm}{
\psfig{file=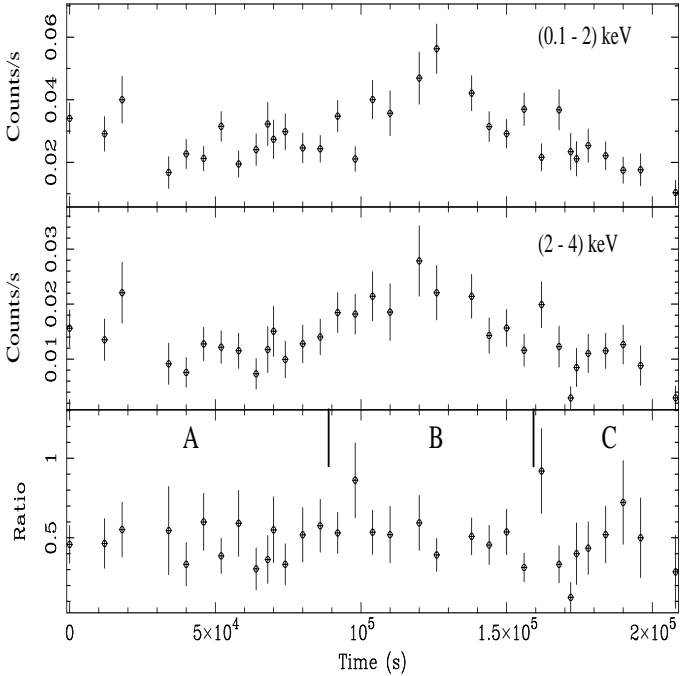,width=9cm,height=9cm,angle=0} }
\caption[]{The $BeppoSAX$ LECS light curves and ratios 
of NGC 7679 in the 
0.1--2 and 2--4 keV energy bands.  
Time bins of 2000 seconds have been used.
The time intervals A, B, and C are those considered to investigate the 
spectral properties of NGC 7679 as a function of the source flux 
(see section 5.1).
} 
\end{figure}

\begin{figure}[htb]
\parbox{10cm}{
\psfig{file=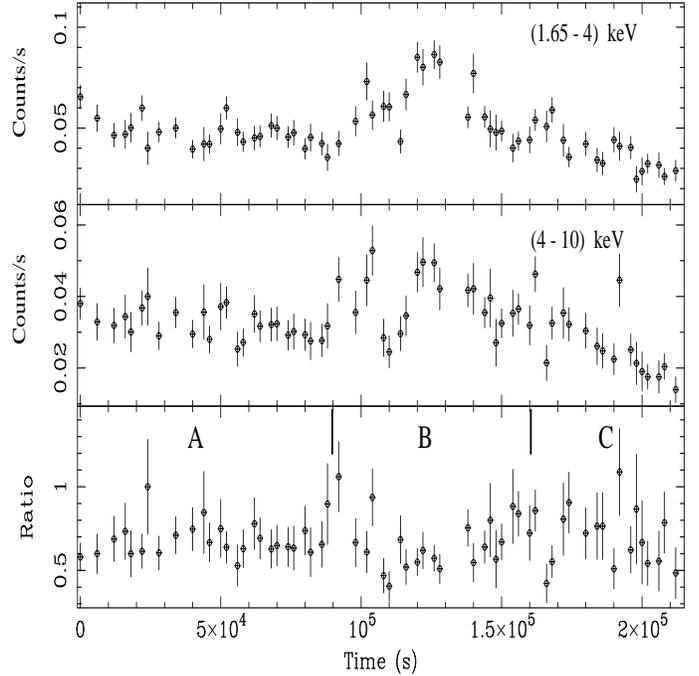,width=9cm,height=9cm,angle=0} }
\ \hspace{0cm} \
\caption[]{The $BeppoSAX$ MECS light curves and ratios 
of NGC 7679 in the 1.65--4 and 4--10 keV energy bands.  
Time bins of 2000 seconds have been used.
The time intervals A, B, and C are those considered to investigate the 
spectral properties of NGC 7679 as a function of the source flux 
(see section 5.1).
}
\end{figure}

\section {X-ray Spectral Analysis}

Spectral channels corresponding to energies 0.12--4 keV and 1.65--10 keV
have been used respectively for the analysis of the LECS and MECS data
(as suggested by the {\it Beppo}SAX Cookbook, Fiore et al., 1999). 
Spectral channels corresponding to energies 15--50 keV have been used 
for the PDS data. 
LECS and MECS source counts have been rebinned to
have a S/N$>5$ in each energy bin. 
Standard response matrices (September 1997 for MECS and January 2000
for LECS) and standard cross-constants normalizations for the instruments 
(see Fiore et al. 1999) have been used in the fitting procedure.
All the models discussed in this paper
have been filtered by the Galactic absorption 
column density ($N_{{\rm H,Gal}} = 4.54\times 10^{20}$ 
cm$^{-2}$, see Table 1).
Quoted errors on the fit parameters give the
90\% confidence intervals for one interesting parameter ($\Delta
\chi^2=2.71$). Finally, thermal emission is 
modeled with the MEKAL plasma emission code.

\subsection{Results}

In the fitting process we have used only LECS and MECS data, since the PDS
data could be contaminated by the X-ray source located $\sim 19^{\prime}$ 
to the south-west 
of NGC 7679 (see Section 3 and below). 
However, the PDS data have been used to investigate the
behavior of the  NGC 7679 spectrum in the energy range $10 - 50$ keV and to  
check for consistency with the LECS/MECS results. 

The simplest interpretation of the overall LECS and 
MECS data is a single power law model with photon index $\Gamma \sim 1.75$ and
small intrinsic absorption, $N_{H} < 4 \times 10^{20}$ cm$^{-2}$ (see Table
3). No emission lines or absorption edges of statistical significance 
are found superimposed
on the power law continuum; the 90\% upper limits on the Fe K lines equivalent
width  at 6.4 and 6.7 keV are 180 eV and 170 eV, respectively.  From a
statistical point of view ($\chi^{2}_{\nu}$ = 0.93), no further spectral
complexity is required.  

\begin{figure*}[htb]
\parbox{10cm}{
\psfig{file=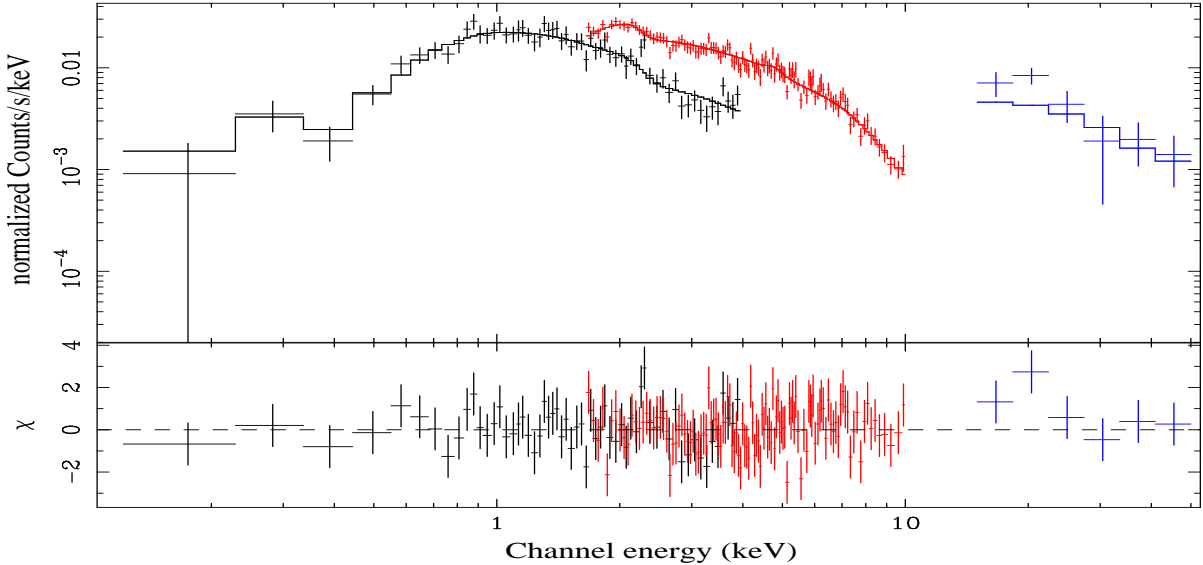,width=16.2cm,height=7.5cm,angle=0} }
\ \hspace{0cm} \
\caption[]{The LECS, MECS and PDS folded spectra of NGC 7679 fitted
with a power law (Table 3). The residuals of the fit are also shown.}
\end{figure*}

The observed spectra and the residuals of the LECS, MECS and PDS
data set compared with the best fit power-law model are reported in Figure 3. 
In the PDS energy range we have also 
included in the model the contribution expected
from the X-ray source detected in the MECS field of view 
(see Section 3), which we estimate to be $\sim 20\%$
\footnote { 
The MECS spectrum of this source is described by a power law model 
with photon index of $1.5^{+0.5}_{-0.4}$. 
The net counts are $245 \pm 23$, corresponding to a
2-10 keV observed flux of $\sim 8\times 10^{-13}$ erg cm$^{-2}$ s$^{-1}$. 
The RASS flux of this object (see section 3) is consistent with 
that estimated using the MECS, 
therefore no sign of variability is present on long timescales.
The PDS regime has been
modeled with the same power law  which  describes the MECS
data; extrapolated to the 10-50 keV range the expected flux 
is $\sim 1.7 \times 10^{-12}$ erg cm$^{-2}$ s$^{-1}$, i.e.  about $20\%$
that from NGC 7679 itself.}.
The very good agreement between the PDS data and the 
extrapolation of the power-law model, which describes the  X-ray spectrum of NGC
7679 in the LECS and MECS energy range, implies that this simple model  can
reproduce the  NGC 7679 spectrum from 0.1 up to 50 keV.
The observed fluxes and intrinsic luminosities of the best fit models
are given in Table 3.
The observed flux is $\sim 6$ times higher than that estimated from the 
$Einstein$ 
IPC observation (0.2-4.0 keV energy range) and about a factor of 4 times 
higher than that measured from the PSPC during the ROSAT All Sky Survey 
(0.1-2.4 keV energy range).

\begin{table}
\caption[]{Results of the spectral analysis. Power law model.}
\begin{flushleft}
\begin{tabular}{  l l l l l l l l l l }
\noalign{\smallskip}
\hline 
\noalign{\smallskip}
\hline
\noalign{\smallskip}
 
 {\it Beppo}SAX LECS and MECS data                      &                    \\
$N_H $ ($10^{20}$ cm$^{-2}$) & 2.2$^{+1.8}_{-1.4}$ \\
$\Gamma$                     & 1.75$^{+0.05}_{-0.05}$ \\
$\chi^2/\nu/\chi^2_{\nu}$    & 160/172/0.93     \\
Flux$^a$ ($10^{-12}$ erg cm$^{-2}$ s$^{-1}$) & 3.3, 6.0, 9.0 \\
$L_x$$^a$  ($10^{42}$  erg s$^{-1}$)           & 8.6, 8.1, 12.2 \\
\hline
\hline
\noalign{\smallskip}

 ASCA  SIS and GIS data                      &                    \\
$N_H $ ($10^{20}$ cm$^{-2}$) & $ < 7.6 $ \\
$\Gamma$                     & 1.72$^{+0.06}_{-0.06}$ \\
$\chi^2/\nu/\chi^2_{\nu}$    & 342/312/1.1     \\
Flux$^a$ ($10^{-12}$ erg cm$^{-2}$ s$^{-1}$) & 2.8, 5.2, 8.3 \\
$L_x$$^a$  ($10^{42}$  erg s$^{-1}$)           & 7.0, 7.1, 11.2 \\
\hline
\hline
\noalign{\smallskip}
\end{tabular} 
\end{flushleft}

\smallskip
$N_H$ is the column density of neutral hydrogen in addition to $N_{H,Gal}$
given in Table 1. 
The LECS/MECS and the SIS/GIS 
normalizations (0.74 and 0.77, respectively)
are consistent with the known differences in
the  absolute calibration  of the instruments. 

$^a$ Observed fluxes and unabsorbed luminosities are given in the 0.1--2, 
2--10, and 10--50 keV bands.
Fluxes and luminosities are relative to the MECS normalization ({\it Beppo}SAX 
data) or to the GIS normalization (ASCA data).
\end{table}

We have also tried a model composed of a soft thermal component plus an 
absorbed power law component.
Besides having been suggested as ``canonical" model for a sample
of low luminosity AGNs, LINERs and starburst galaxies observed with $ASCA$
(Ptak et al. 1999) such a model also describes the case of a starburst
dominating the soft emission and a (heavily obscured) AGN dominating the 
hard one. 
Leaving the abundance as a free parameter, its best fit value is very low 
($ A \sim 0.01$) and is probably unfeasible for a starburst.
Fixing the abundance to the solar ones,
the best fit values are $kT \sim 1.2$ keV
and $\Gamma = 1.72^{+0.07}_{-0.12}$, again with  
small intrinsic absorption ($N_H < 4.5\times 10^{20}$ cm$^{-2}$), 
giving unabsorbed luminosities (0.1--10 keV) 
of the  thermal and power law components 
of $3.8 \times 10^{41}$ \es and $1.6 \times 10^{43}$ \es, respectively. 
Such a soft thermal component could be accommodated within the data 
uncertainties but it is not
statistically required by the F test ($ \Delta \chi^2$ = 1 for 2 
additional parameters).

Finally, in order to study in detail the spectral properties of NGC 7679 as the
flux level changes, we repeated the spectral analysis using the 
LECS and MECS counts
\footnote {
Due to the limited statistics available in the PDS, it was not possible 
to split the observation in 3 segments. 
The PDS data will not be considered here.
}
from the three different time intervals shown in figure 1 and 2: 
A (quiescent status), B (flux increase)  
and C (flux decrease). 

Each spectrum is well fitted by a single power law
model with spectral parameters statistically consistent with 
those  of the total spectrum (see Table 3). This confirms our expectations based on
the  constancy of the ratios of the count rates in various
energy bands (see Sect. 4). No thermal components are required.
There is  some marginal evidence for the presence of line-like residuals 
around 7.1 keV when  NGC 7679 is brighter (time interval B) , and  
around 6.2 keV  when NGC 7679 is weaker (time interval C). 

\section {ASCA Observations}

NGC 7679 was observed by the ASCA satellite (Tanaka, Inoue and Holt, 1994)
on July 1, 1998. 
The data were retrieved from the ASCA archive 
and were cleaned by applying the standard selection criteria described 
in the TARTARUS database (see http://tartarus.gsfc.nasa.gov)
and using the same software packages used for the {\it Beppo}SAX data.
The journal of the observation and source count rates are given in Table 4.

\begin{table*}
\caption[Table 2.]{ASCA Observation Log}
\begin{flushleft}
\begin{tabular}{llllllllllllll}
\noalign{\smallskip}
\hline
\noalign{\smallskip}
Sequence & Date          & \multicolumn{2}{l}{Exposure time (ks)}  &
\multicolumn{2}{l}{Count Rate$^a$ (counts s$^{-1}$)}                 \\
\cline{3-4}\cline{5-6}
 Number &     & SIS01 & GIS23 & SIS01 & GIS23  \\
        &     &       &       & 1 -- 10 keV & 0.6 -- 10 keV \\
\noalign{\smallskip}
\hline
\noalign{\smallskip}
66019000 & 1998 July 1 & 88.8             & 98.3 &  
                         0.088$\pm 0.001$   & 0.079 $\pm 0.001$     \\ 
\noalign{\smallskip}
\hline
\end{tabular} 
\end{flushleft}
\bigskip

$^a$ Background subtracted source count rates,  
with $1 \sigma$  photon
counting statistics errors.  Extraction radii for SIS and GIS are 
$4^{\prime}$ and $6^{\prime}$,  respectively.  The net counts from NGC 7679
represent about $81\%$ of the total gross counts in the GIS or SIS 
source region.

\end{table*}

\subsection {Data Preparation}           
  
Total counts (source + background)  were
extracted from a circular region of 4$^{\prime}$ 
radius for SIS0 and SIS1 and a 6$^{\prime}$ radius for GIS2 and GIS3. 
For the SIS,
background counts were taken from the remaining area of the chip, while for the
GIS, background counts were taken from source-free circular  regions close to
NGC 7679. 
We exclude from the spectral analysis all SIS data with E $<$ 1.0 keV  
and all the GIS data with E $<$ 0.6 keV 
(see http://heasarc.gsfc.nasa.gov/docs/asca/watchout.html).
No other X-ray sources are detected in the SIS or GIS field 
of view (note that the X-ray source detected in the MECS image, 
at about $19^{\prime}$  to the south-west of NGC 7679, 
is outside the ASCA GIS or SIS field of view). 

For the GIS data we use the detector Redistribution Matrix Files (RMF)
gis2v4$\_$0.rmf and gis3v4$\_$0.rmf, respectively; for the SIS data the
RMF files were obtained using the FTOOLS task SISRMG.  
The Ancillary Response Files (ARF) for both SIS and GIS 
were created with the FTOOLS task ASCAARF at the location of NGC
7679 in the detectors.

In order to improve on the statistics we have produced a combined SIS spectrum
(SIS01 hereafter) and a combined GIS spectrum
(GIS23 hereafter) along with the relative background files and calibrations
(see http://heasarc.gsfc.nasa.gov/docs/asca/abc/abc.html). 
For the spectral analysis, source counts have been
rebinned to have a S/N$>5$ in each energy bin.  

\subsection {ASCA Results} 

The ASCA SIS and GIS light curves show a behavior very similar to those
obtained using the {\it Beppo}SAX instruments (see Figure 1 and 2), i.e. 
a variation of the count rates of a factor of 2-3 over the total 
observation, and a similar minimum doubling/halving time.  

The ASCA (SIS01, GIS23) spectrum of NGC 7679 over the whole observing  period
is well described by a power law model with 
$\Gamma \sim 1.72$ and small intrinsic absorption, 
$N_{H} < 7.6 \times 10^{20}$ cm$^{-2}$ (see Table 3 and Figure 4).
More complex models are not required statistically.
As for the {\it Beppo}SAX data, 
no spectral variations were found as a function of the source count rate, 
as expected from the constancy (within the
errors) of the ratios of the count rates in various energy  bands (e.g., 0.6
- 2 keV vs 2-10 keV). 
There is some evidence  for an FeK line in the ASCA spectrum. 
Over the whole observing  period, this line is located around  6.6
keV (rest frame); in agreement with the {\it BeppoSAX} results the line best-fit energy
also appears to be higher ($\sim 6.9$ keV) when the  source is brighter, and lower 
($\sim 6.3$ keV) when the source is
weaker. However, the present statistics, either from {\it Beppo}SAX or ASCA, 
are too poor to be conclusive and require further observations by, e.g., XMM
or {\it Chandra}.
We note here only that such variability of the properties of the Iron lines 
has been found to be very common in a sample of Seyfert 1 galaxies 
investigated with ASCA (Weaver, Gelbord and Yaqoob, 2001).

\begin{figure*}[htb]
\parbox{10cm}{
\psfig{file=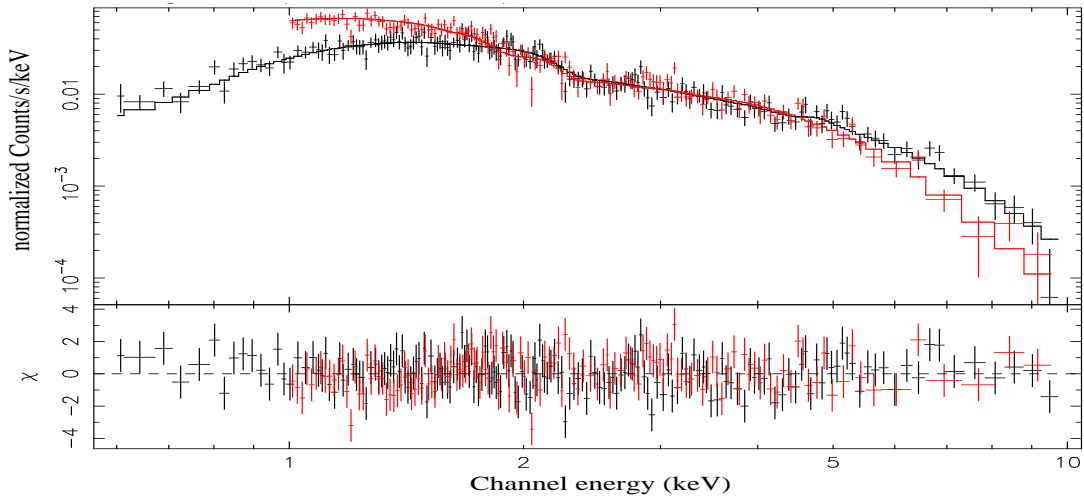,width=16.2cm,height=7.5cm,angle=0} }
\ \hspace{0cm} \
\caption[]{The ASCA SIS01 and GIS23 folded spectra of NGC 7679 fitted
with a power law model (see Table 3).
The residuals of the fit are also shown.}
\end{figure*}

\section {Discussion}

\subsection {Starburst or AGN?}

All the X-ray (0.1--50 keV) properties presented here (Section 4:
variability; Section 5 and 6: spectra)  provide strong evidence in favor of
the existence of an AGN in NGC 7679. 
Indeed, as summarized below, the presence of an AGN is overwhelming
in X rays with respect to that of a starburst. 

The minimum detectable doubling/halving variability time,  $t_{var} \sim$ 15
ksec, implies (using light crossing arguments) an emitting region 
with a size $\leq \sim 4.5\times 10^{14}$ cm  ($1.5\times
10^{-4}$ pc), much smaller than the starburst size ($\sim 10$ Kpc).  
Furthermore, the maximum observed variation,
$\Delta L_{2-10 keV} \simeq 5 \times 10^{42}$ erg s$^{-1}$,
is very high for being produced inside a starburst nucleus.

A simple power-law spectral model ($\Gamma \sim 1.75$) with small intrinsic
absorption ($N_H < 4 \times 10^{20}$ cm$^{-2}$)  provides a very good
description of the spectral properties of NGC 7679 from 0.1 up to 50 keV.  
This spectral shape along  with the variability properties closely 
resembles that
typically observed in Seyfert 1 galaxies (Nandra et al., 1997). 

Finally, the total X-ray luminosity of NGC 7679 ($L_{0.1-10\ keV} \sim 1.7\times
10^{43}$ erg s$^{-1}$) also places it closer to AGNs than to starbursts.  The
most X-ray luminous star-forming galaxy yet detected, NGC3256 (Moran et al.
1999),  has $L_{0.1-10 keV} \simeq 2.2\times 10^{42}$ erg s$^{-1}$, which is $\sim 8$
times weaker than that of NGC 7679.  Note instead that the infrared luminosity
of NGC3256 ($L_{FIR} \simeq 6 \times 10^{11}$ $L_{\odot}$)
is $\sim 5$ times higher than that of NGC 7679, and so the starburst
is likely to be more powerful in NGC3256.

What about the possible starburst contribution to the X-ray emission?
In Section 5.1 we have shown that a soft thermal component can be
accommodated within the data uncertainties;
its (0.1 - 10 keV) luminosity ($\simeq 3.8\times 10^{41}$ erg
s$^{-1}$) is $\sim$ 5 times lower than the total thermal emission in NGC3256, 
in better accordance with the lower infrared luminosity of NGC 7679.

\subsection {Which kind of AGN ?}

Given that the bulk of
the  X-ray emission is powered by an AGN, in this section we try to 
specify  which kind of AGN better accounts for the observed X-ray 
properties of NGC 7679.

The X-ray spectrum does not show any evidence of the
characteristic cut-off  at low energies and of the strong Iron-K line(s) at
$\sim 6.4-6.9$ keV  usually found in absorbed Type 2 Seyferts with $N_H \sim
10^{23} - 10^{24}$ cm$^{-2}$ (see Bassani et al., 1999 and references therein).
Thus a Compton thin Seyfert 2 could be excluded.

One interesting possibility could be a Compton thick ($N_H \gae 5\times 10^{24}$ cm$^{-2}$)
Seyfert 2 galaxy.  In this case the observed X-ray spectrum should be totally
dominated  by scattered radiation from an otherwise invisible nucleus.  
Typical examples of this class of ``Compton Thick" sources 
are NGC 6240 (Vignati et al., 1999) and NGC 4945 (Guainazzi et al., 2000a).
However the observed X-ray properties suggest against this hypothesis. First, we
do not see the very strong (EW $\sim 1-2$ keV) Iron-K line usually observed in 
``reflection dominated" Seyfert 2 (Matt et al., 2000; Bassani et al., 1999).  
Second, as pointed out by Guainazzi et al. (2000b) 
in the case of IRAS12393+3520, the observed variability 
introduces severe constraints on this possibility  
since for $t_{var} \sim 15$ ksec
the scattering/absorption should occur very close to the nucleus ($\sim$
a few $\times 10^{-4}$ pc), in net contrast with the idea that the matter
obscuring the  nuclear region is located at a distance of 1-100 pc.  
Finally, the good agreement between the LECS/MECS spectra with the PDS data up
to 50 keV excludes the presence of any absorption cut-off which is  
expected around 20 keV for Compton-thick objects.

To conclude, the only kind of AGN consistent with the
X-ray properties of NGC 7679 is a  Seyfert 1 AGN.
However, its broad-band properties make NGC 7679  
a peculiar Seyfert 1 galaxy. 
First of all, the detection of a broad H$\alpha$ component without 
a broad H$\beta$ component is more typical of Type 1.9 Seyferts, which 
are usually characterized by X-ray absorbed spectra (see Bassani et al.,1999 and 
references therein ).
Second, NGC 7679 is underluminous in the
optical emission lines with respect to ``normal"  Seyfert 1. Using the 
$L_{(2-10 keV)}$ - $L_{H \alpha}$ relationship observed for Seyfert 1's and QSO
(see Maiolino et al., 2001) and given the  observed $L_{(2-10 keV)} \sim 8
\times 10^{42}$ \es, we would expect an  $H \alpha$  luminosity in the range ($1
\sigma$)  $\sim 4 \times 10^{41}$  -- $\sim 6 \times 10^{42}$ \es. This is at
least a factor of 5 higher than the observed  broad $H \alpha$ luminosity 
($\sim 8 \times 10^{40}$ erg s$^{-1}$, Keweley, private communication) in 
NGC 7679. 

\subsection {Implications}

If we combine the results presented here (i.e. clear predominance of 
an AGN in the X-ray regime) together with the clear predominance 
of a starburst in the optical and IR regime (see Section 2) we 
conclude that we have discovered another ``Composite" galaxy.  

Although Starburst/Seyfert composites have been studied in the last two  
decades, we find here that the broad-band properties of NGC 7679 
are different from most of the composite galaxies studied so far.
Composite Starburst/Seyfert galaxies were first  discovered and discussed  by
Veron et al. (1981a) and Veron, Veron and Zuiderwijk (1981b); similar objects
were also found in the EMSS  (Halpern et al. 1995), among serendipitously
detected IR-bright $Einstein$ sources (Moran et al. 1996a) and in deep $ROSAT$
fields (Boyle et al. 1995, Griffiths et al., 1996).  
Composite Seyfert 2 / Starburst galaxies have been recently studied  by
Levenson, Weaver and Heckman, 2001 and the AGN / starburst inter-connection   has
been also used to explain the overall spectral properties of the ULIRG   
(Iwasawa, 1999).  
However, in all these objects it is now clear  that heavy absorption in the
nuclear region and a strong  circumnuclear starburst are responsible  for their
broad-band  properties (e.g. infrared, optical and X-ray regime, see. e.g. 
Iwasawa, 1999; Levenson, Weaver and Heckman, 2001).  
In particular, much of the observed far-infrared emission 
may be associated with the starburst, which also 
can contribute significantly  to the observed X-ray flux in the soft 
(E $ < $ few keV) energy band; the nuclear X-ray source is 
generally absorbed with $N_H > 10^{23}$ cm$^{-2}$. 
In these objects the circumnuclear starburst should 
also play a significant role in the obscuration process 
(Levenson, Weaver and Heckman, 2001, Levenson et al., 2001).
The most striking peculiarity of NGC 7679, at odds with the
above examples, is that {\it its X-ray  spectrum is not highly absorbed 
($N_H < 4\times 10^{20}$  cm$^{-2}$) and strong (EW $\sim$ 1 keV) 
Iron lines (which could  suggest a Compton-thick Type 2 AGN) are not present.}

Have other galaxies with X-ray, optical and infrared properties similar to 
NGC 7679 been detected before ? Six
\footnote{
One  of the objects  in Moran's list (IRAS10113+1736)
is no longer a valid 
candidate since subsequent observations of all the 7 objects in the original
sample have shown that, for this object, the IR and the X-ray emission 
originate from different sources (Condon et al., 1998).}
low redshift galaxies, with optical spectra dominated by the  features of
HII galaxies and by the presence of weak and elusive 
$H \alpha$ broad wings (probably indicative of a faint or obscured AGN)
but X-ray luminosities typical of broad line AGNs (i.e. $L_x$
ranging from $1.5\times 10^{42}$ erg s$^{-1}$ to $5\times 10^{43}$ erg s$^{-1}$
in the $ROSAT$ band)  have been
singled out in a large spectroscopic optical survey of 210 bright IRAS and X-ray
selected sources (Moran et al. 1996b).  The L(0.1--2.4 keV)/L$_{FIR}$ and the
$O[III] \over H \beta$ flux ratio  of NGC 7679  are in agreement with the
values shown by these starburst/Seyfert composites.
Only one of the objects in the Moran et al., 1996b list 
has been studied in detail (IRAS00317-2142; Georgantopoulos, 2000) 
using ASCA data, while 
for the remaining 5 objects only low signal-to-noise ROSAT All Sky Survey data 
are available at present and no spectral or variability X-ray 
properties are known to date.
IRAS00317-2142 has clear Seyfert 1 properties in the X-ray domain; in particular, 
it has a low absorption column density and no other signs of possible obscuration 
(e.g. strong Iron lines). 
It is therefore very similar to NGC 7679 in many aspects.
Another object with overall properties very similar to 
NGC 7679 (but not included in Moran's list) 
could be IRAS12393+3520, which has been investigated in detail 
by Guainazzi et al., 2000b using PSPC and ASCA data.
The interpretation for these objects is
that the optical spectra are influenced heavily by the starburst component,
while the Seyfert component is responsible for most of the intense X-ray
emission. 
As pointed out by Moran et al. (1996b), their peculiarity is not the strength of their starburst but the
apparent optical weakness of the AGN.
It is possible
that a dusty ionized absorber is able to selectively obscure the optical 
emission, leaving the X-ray emission almost unabsorbed (Georgantopoulos, 2000; Maiolino et al. 2001).
However, the physical nature of these objects, 
the reconciliation of their optical and X-ray spectra, their 
role inside the Unification Scheme of AGN and the understanding of how the 
absorption process works in these objects and 
modifies their global properties are far from obvious and are still unknown.

\section {Conclusions}

On the basis of the spectral shape, the high luminosity, and 
the observed X-ray variability, we conclude that the X-ray 
emission of NGC 7679 is strongly dominated by the presence of an AGN.  
The starburst activity, revealed
by the IR emission, by optical spectroscopy as well as by H$ \alpha$ imaging,
 and dominating the optical and IR bands, is clearly 
overwhelmed by the AGN in the X-ray band. 
At first glance, this is similar to what is observed in other composite starburst/AGN galaxies, e.g. 
NGC 6240 (Vignati et al., 1999) and NGC 4945 (Guainazzi et al., 2000b).
However, at odds with the above examples, the X-ray 
spectrum of NGC 7679 does not show any signs of absorption.

What are the most important implications of this result?
Objects like NGC 7679 show that the absorption process 
can work under very different scenarios.
In this object we have unabsorbed X-ray emission but possible absorbed 
(or intrinsically weak) optical emission from an AGN.
This scenario must be understood and included in the 
general framework of the AGN Unification Scheme, 
along with the usual Type 1 and Type 2  AGN, the 
Broad absorption line QSOs etc.. 
(see Gilli et al., 2001 for an observational and updated summary  
on how the absorption process modifies the optical and X-ray properties 
in the AGN family).
From an observational point of view, 
optical and infrared spectroscopy could be
highly inefficient in revealing the presence of an AGN in objects like 
NGC 7679, 
which is unambiguously  revealed by X-ray spectroscopic and 
variability  investigations.  
This consideration has strong implications for the  
optical identification work of the faint and low S/N X-ray sources 
in the XMM and Chandra fields; in this case the X-ray data are usually not
good enough to perform  a detailed X-ray investigation and the optical and
infrared  observations could misidentify the 
real source of the X-ray emission. 
Finally, the observed properties of NGC 7679 
(e.g. the existence of a bar, the signs of tidal distortion with a nearby faint companion, etc.. 
see Section 2)  could make it a low-redshift 
example of the 
peculiar and interacting galaxies found by deep HST-NICMOS observations. Their
number density seems to increase with $z$, being $\sim 4\%$ of all galaxies 
locally  and $\sim
14\%$ at z$ \sim 1$ (see Corbin et al., 2000 and references therein).
We believe that the new generation of X-ray satellites, 
like XMM-Newton and {\it Chandra}, will
find  many sources similar to NGC 7679 and will help us 
investigate  their nature in detail.

Indeed galaxies with bright X-ray emission ($L_x \sim 10^{42-43}$ \es)  but
weak or absent AGN features in the optical band have been known  since {\it
Einstein} observations (Elvis et al., 1981; Tananbaum et al., 1997) and are now
showing up  in Chandra deep-fields  (e.g. Hornschemeier et al., 2001; Fiore et
al., 2000). Although these sources usually lack evidence for strong starburst
emission and, therefore, their nuclei are not usually selected for more
detailed  and deeper optical studies, the main problem is the same as for
objects like NGC 7679, i.e. to explain the optical weakness/disappearance of
the AGN in these X-ray  luminous sources. 

\begin{acknowledgements}

RDC wish to thank ESO in Garching for the hospitality during the 
preparation of part of this work.
We thank A.Guarnieri and V.Zitelli for the optical spectroscopic 
observation of NGC 7679 at the Loiano Telescope, L. Kewley for 
having given  us many results on NGC 7679 in advance of publication
and A. Caccianiga, A. Comastri and P. Severgnini for  useful discussions.
This research has made use of 
SAXDAS linearized and cleaned event
files produced at the {\it Beppo}SAX Science Data Center,  
the ASCA archive,
the TARTARUS database.
and the NASA/IPAC extragalactic database (NED).
We thank all the members working at the above mentioned projects. 
This work received partial financial support from ASI (ARS-9974) 
and MURST.

\end{acknowledgements}

\end{document}